\documentclass[aps,showpacs,superscriptaddress,preprint]{revtex4}%
\usepackage{amsfonts}
\usepackage{amsmath}
\usepackage{amssymb}
\usepackage{graphicx}%
\providecommand{\U}[1]{\protect\rule{.1in}{.1in}}

\begin{document}
\title{Antifferomagnetic FeSe monolayer on SiTiO$_{3}$: The charge doping and
electric field effects}
\author{Fawei Zheng}
\affiliation{LCP, Institute of Applied Physics and Computational Mathematics, Beijing
100088, People's Republic of China}
\affiliation{Beijing Computational Science Research Center, Beijing 100084, People's
Republic of China}
\author{Zhigang Wang}
\affiliation{LCP, Institute of Applied Physics and Computational Mathematics, Beijing
100088, People's Republic of China}
\affiliation{Beijing Computational Science Research Center, Beijing 100084, People's
Republic of China}
\author{Wei Kang}
\affiliation{Center for Applied Physics and Technology, Peking University, Beijing 100871,
People's Republic of China}
\author{Ping Zhang}
\thanks{Author to whom correspondence should be addressed. E-mail: zhang\_ping@iapcm.ac.cn}
\affiliation{LCP, Institute of Applied Physics and Computational Mathematics, Beijing
100088, People's Republic of China}
\affiliation{Beijing Computational Science Research Center, Beijing 100084, People's
Republic of China}
\affiliation{Center for Applied Physics and Technology, Peking University, Beijing 100871,
People's Republic of China}


\begin{abstract}
We present theoretically the electronic structure of
antiferromagnetic (AFM) FeSe monolayer on TiO$_{2}$ terminated
SrTiO$_{3}$(001) surface. It is revealed that the striking
disappearance of the Fermi surface around the Brillouin zone (BZ)
center can be well explained by the antiferromatnetic (AFM) phase.
We show that the system has a considerable charge transfer from
SrTiO$_{3}$(001) substrate to FeSe monolayer, and so has a
self-constructed electric field. The FeSe monolayer band structure
near the BZ center is sensitive to charge doping, and the
spin-resolved energy bands at BZ corner are distorted to be
flattened by the perpendicular electric field. We propose a
tight-binding model Hamiltonian to take these key factors into
account. We also show that this composite structure is an ideal
electron-hole bilayer system, with electrons and holes respectively
formed in FeSe monolayer and TiO$_{2}$ surface layer.

\end{abstract}
\maketitle

Due to its simple crystal structure, prominent antiferromagnetic magnetic
ordering, and significant pressure effect on superconductivity, during the
last few years FeSe binary superconductor has been drawing extensive attention
from fundamental mechanism of Fe-based superconductivity to promising
application in such areas as superconducting wires and thin films \cite{Wu2008}. Recently,
monolayer FeSe with atomic flatness was successfully grown on SrTiO$_{3}$(001)
surface \cite{Xue2012}, and strikingly, the high-temperature
superconductivity signature from the monolayer was revealed by high energy
resolution scanning tunneling spectroscopy (STS). It is even hopeful that with
further improving the sample quality, the transition temperature is probably
as high as 77 K, the liquid nitrogen boiling point. The subsequent
angle-resolved photoemission spectroscopy (ARPES) experiment \cite{Zhou2012}
shows that the Fermi surface of this FeSe monolayer only consists of electron
pockets near the Brillouin zone (BZ) corner, with no hole pockets around the
zone center observed. Theoretical efforts are being paid to explain these
intriguing STS and ARPES reports. The soft phonon of SrTiO$_{3}$(001) was
found to strengthen the Cooper pairing \cite{Lee2012}. $S_{4}$ symmetry was
also noticed \cite{Hu2012}, wherein it was suggested that cuprates and
iron-based superconductors share an identical high-T$_{c}$ superconducting
mechanism. The density functional theory (DFT) total-energy calculations gave
that collinear antiferromagnetic order (CAFM) FeSe monolayer on TiO$_{2}$
terminated SrTiO$_{3}$(001) surface is the most stable structure, and there is
neither hybridization nor charge transfer between FeSe monolayer and
SrTiO$_{3}$(001) surface \cite{Lu2012}. The calculated Fermi surface for this
CAFM structure exists around the M point in the fold BZ, and there
is a Dirac-cone-like bands of FeSe near Fermi energy, which unfortunately are
not observed by the ARPES measurement \cite{Zhou2012} that the Fermi surface
disappears around the zone center. A more recent DFT study \cite{Kohn} on a single FeSe
monolayer (without substrate) showed that an AFM instead of CAFM phase can have the Fermi surface
resembling the ARPES data. Basically, however, the interaction between
AFM-ordered FeSe monolayer and SrTiO$_{3}$(001) surface remains largely
unclear in literature, despite its urgent importance on understanding the key
role played by the interface in controlling the Fermi surface topology and
high-temperature superconductivity of FeSe monolayer.

In this work, we carefully investigate with DFT based \textit{ab initio}
calculations the electronic structure of AFM FeSe monolayer on TiO$_{2}$
terminated SrTiO$_{3}$(001) surface. We show that the calculated electronic
structure of the system is close to the ARPES experimental results. The charge
transfer and electric field are induced by the interaction between SrTiO$_{3}%
$(001) substrate and AFM FeSe monolayer, and turn back out to play an
important role in modulating the band structure and Fermi-surface topology of
the FeSe monolayer. One extraordinary consequence is that the spin-resolved
electron pockets surrounding the BZ corner are distorted and become flattened.
We propose a tight-binding model Hamiltonian that can describe these key factors.

\begin{figure}[ptb]
\begin{center}
\includegraphics[width=0.4\linewidth]{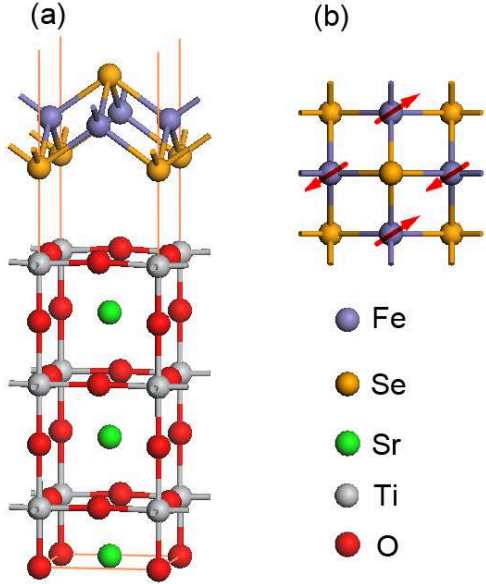}
\end{center}
\caption{(Color online) The side view (a) and top view (b) for the atomic
structure of monolayer FeSe absorbed on TiO$_{2}$ terminated SrTiO$_{3}$(001)
surface. The magnetic ordering is also represented in top view (b) as red
arrows.}%
\label{structure}%
\end{figure}

We use 6-layer SrTiO$_{3}$(001) slabs to mimic the surface of SrTiO$_{3}$
substrate, as shown in Fig. \ref{structure}(a). The SrTiO$_{3}$(001) surface
is either SrO or TiO$_{2}$ terminated with the nearly same surface energy
\cite{STOsurface1,STOsurface2,STOsurface3}. According to the experiment
\cite{Xue2012}, we use the TiO$_{2}$ terminated surface in our atomic models.
FeSe monolayer is adsorbed on the 1$\times$1 two-dimensional unit cell of
TiO$_{2}$-terminated SrTiO$_{3}$(001) surface. To test the influence of
SrTiO$_{3}$ antiferrodistortion on surface electronic structures, we also
calculated a $\sqrt{2}\times\sqrt{2}$ supercell, in which the lower four
layers of TiO$_{2}$ and SrO are fixed to their bulk positions and the
octahedral rotation angle is set to the experimental value 2.1$^{\circ}$
\cite{STOdistortion}. In our models, the vacuum space is larger than 10 \AA .
The results are also checked for vacuum space larger than 20 \AA . The bulk
calculation shows that the lattice parameter of SrTiO$_{3}$ is 3.901 \AA ,
which agrees well with experiments and previous theoretical results. The
adsorption structure relaxation is performed with fixed lattice parameters.

The total energy and electronic structure calculations are performed
by using projector augmented wave (PAW) method \cite{PAW1,PAW2}. The
exchange correlation potential is described by the generalized
gradient approximation (GGA) of Perdew-Burke-Ernzerhof (PBE) type
\cite{PBE}. The plane wave cut-off energy is chosen to be 400 eV,
which is converged in our test. 9$\times $9$\times$1 and
6$\times$6$\times$1 Monkhorst-Pack $k$-points are used in the
reciprocal space integration for 1$\times$1 and
$\sqrt{2}\times\sqrt{2}$ supercells. The system is relaxed until the
force on each atom is smaller than 0.01 eV/\r{A}. To include the
strong correlation effect for \textit{d} electrons of Fe atoms, we
adopt Hubbard-$U$ correction. We use $U$=0.5 eV for Fe-$3d$ electrons in our main calculations, and the
influence of different values of Hubbard-$U$ on electronic structures
are also discussed in the following text. All the DFT calculations
are performed by using Vienna Ab-initio Simulation Package (VASP)
\cite{Kresse1996}. The relaxed adsorption structure of FeSe
monolayer on TiO$_{2}$ terminated SrTiO$_{3}$(001) surface is shown
in Fig. \ref{structure}. The lower layer Se atoms are on the top of
Ti atoms, while the Fe atoms locate on the top of O atoms. The
vertical distance between Ti and Se atoms is 3.13 \r{A}, which is
slightly larger than that of the CAFM phase by 0.07 \r{A}. The
vertical distance between Fe and O atoms is 4.43 \r{A}. The O atoms
in the TiO$_{2}$ top-layer arise slightly, they are 0.1
\r{A}\thinspace\ higher than Ti atoms. The magnetic moment on each
atom is calculated by integrating the spin density in Wigner-Seitz
radius of 1.30 \r{A}, 1.16 \r{A}, 2.14 \r{A}, 1.22 \r{A}, and 0.82
\r{A}\thinspace\ for Fe, Se, Sr, Ti, and O atoms respectively. The
calculated magnetic moment is 2.4 $\mu_{B}$ on each Fe atom and is
negligible on the other atoms.

\begin{figure}[ptb]
\begin{center}
\includegraphics[width=0.6\linewidth]{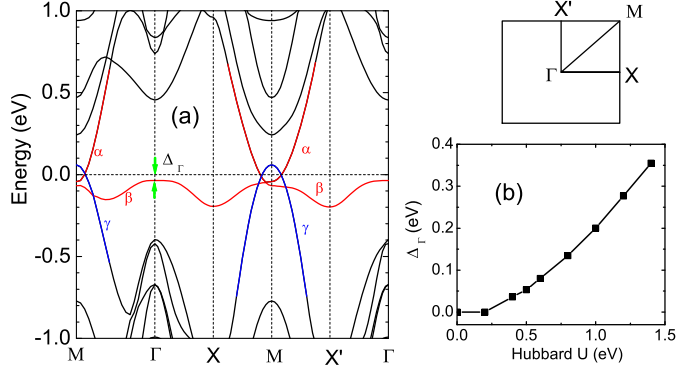}
\end{center}
\caption{(Color online) The spin-up energy bands of monolayer FeSe
adsorbed on TiO$_{2}$ terminated SrTiO$_{3}$(001) surface (a). The
red and blue lines are contributed by 3\textit{d} orbitals on Fe
atoms and 2\textit{p} orbitals on O atoms, respectively. The Fermi
level is set to be zero. Panel (b) shows the
influence of Hubbard-$U$ on the energy difference $\Delta_{\Gamma}$.}%
\label{band}%
\end{figure}

The energy bands for spin-up channel are shown in Fig.
\ref{band}(a). The spin-down energy bands are the same as those for
spin-up channel except for an exchange of X and X$^{\prime}$ points
in BZ. There are two bands [indicated as $\alpha$ and $\gamma$ in
Fig. \ref{band}(a)] crossing the Fermi level $E_{F}$ near M point
and foming an ideal pair of electron-hole bands with perfect
Fermi-surface nesting, while around $\Gamma$ point no metallic bands
occur. Thus, the calculated Fermi surface only exists at the BZ
corner, which agrees with the ARPES result in experiment
\cite{Zhou2012}. In our calculations, the antiferrodistortion of
SrTiO$_{3}$ substrate does not have considerable influence to the
electronic structure of adsorbed AFM FeSe monolayer.

The disappearance of Fermi surface at BZ center is the result of a
full electron filling of band $\beta$. We find that the shape of
band $\beta$ can be altered dramatically by changing the value of
Hubbard-$U$ correction. Actually, band $\beta$ shall cross the Fermi
level at the BZ center in the absence of Hubbard-$U$ correction.
When the Hubbard-$U$ correction is applied, band $\beta$ lowers its
energy and a gap of 41 meV [indicated by green arrows in Fig.
\ref{band}(a)] is obtained for $U\mathtt{=}$0.5 eV, which agrees
with the experimental result \cite{Zhou2012}. The detailed
relationship between the minimum energy difference of band $\beta$
and $E_{F}$ around the BZ center (denoted as $\Delta_{\Gamma}$) and
the Hubbard-$U$ correction is shown in Fig. \ref{band}(b). One can
see that $\Delta_{\Gamma}$ and $U$ are mainly in a linear relation.
We have checked that this phenomenon does not exist in the
paramagnetic FeSe/SrTiO$_{3}$ system, wherein there are energy bands
crossing the Fermi level around the BZ center even for 5.0 eV
Hubbard-$U$ correction.

Our PDOS (partial density of states) analysis shows that bands
$\alpha$ and $\beta$ are contributed by Fe $3d$ orbitals, while the
hole-type band $\gamma$ comes from the O $2p$ orbitals. We analyze
the wavefunctions of band $\beta$ at $\Gamma$ point and bands
$\alpha$, $\beta$ and $\gamma$ at M point. The wavefunction of band
$\beta$ at $\Gamma$ point shows a shape of a dumbbell in
$z$-direction with a torus in $xy$-plane, which is the typical
$d_{z^{2}}$ feature. Whereas, the wavefunctions of bands $\alpha$
and $\beta$ at M point show vertical four-leaf shapes, which are the
$d_{xz}$ and $d_{yz}$ features. The wavefunctions for band $\gamma$
are dumbbell shaped $p_{x}$ and $p_{y}$ orbitals that are mainly
distributed on the surface O atoms. The crossing between bands
$\alpha$ and $\gamma$ indicates that charge transfer occurs from
surface O atoms of SrTiO$_{3}$ substrate to Fe atoms of FeSe
monolayer. Our Bader analysis identifies a charge transfer of
$-$0.17 $e$ per unit cell.

\begin{figure}[ptb]
\begin{center}
\includegraphics[width=0.5\linewidth]{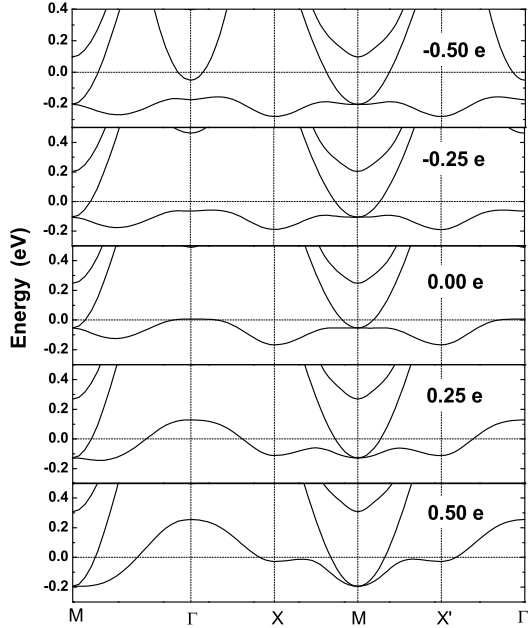}
\end{center}
\caption{(Color online) The electronic energy bands of free-standing
monolayer FeSe with different levels of charge doping. The value of
charge doping is changed from $-0.5$ $e$ to 0.5 $e$ per unit cell. The Fermi energy is set to zero.}%
\label{band_charge_doping}%
\end{figure}

From above results, we know that the AFM FeSe monolayer is charge
doped by the SrTiO$_{3}$ substrate. In order to analyze this effect,
we study a free-standing AFM FeSe monolayer by doping charge from
$-$0.5 $e$ to 0.5 $e$. The results are shown in Fig.
\ref{band_charge_doping}. We see that the shape of band $\beta$
changes dramatically when the system is positively charged. Band
$\beta$ at $\Gamma$ point crosses the Fermi level, and becomes a
hole pocket. When the FeSe monolayer is negatively charged, the
shape of band $\beta$ keeps static, while band $\alpha$ at $\Gamma$
point lowers its energy and finally crosses the Fermi energy,
showing an electron pocket. Both positive and negative charge doping
increase $E_{F}$ relative to the energy of band $\alpha$ and $\beta$
at M point. These results agree well with the previous study on FeSe
monolayer without Hubbard-$U$ correction \cite{Kohn}. However, as we
have determined above, the realistic charge transfer in
FeSe/SrTiO$_{3}$ is only $-0.17e$. By comparison with Fig.
\ref{band_charge_doping}, we find that this value is not large
enough for the existence of electron pocket at $\Gamma$ point. Thus,
the energy bands of AFM FeSe monolayer are altered limitedly by pure
charge doping from SrTiO$_{3}$ substrate.

\begin{figure}[ptb]
\begin{center}
\includegraphics[width=0.5\linewidth]{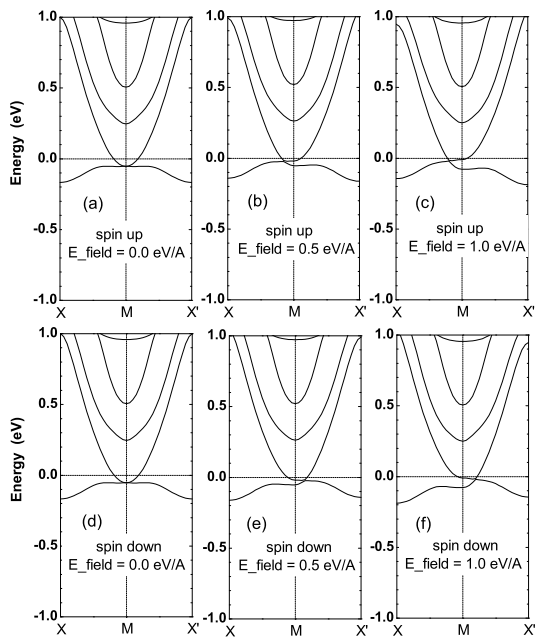}
\end{center}
\caption{(Color online) The electronic energy bands of free-standing
monolayer FeSe with different vertical electric field strength. The
value of electric field strength is changed from 0.0 eV/\AA \,\,to
1.0 eV/\AA . The energy bands near Fermi energy are distorted and
show different behaviors for spin-up and spin-down channels.
The Fermi energy is set to zero.}%
\label{Efield_DFT}%
\end{figure}

Besides the charge doping, another prominnet feature in Fig.
\ref{band} is that bands $\alpha$ and $\beta$ are nonsymmetric along
M-X and M-X$^{\prime}$ directions, and a small splitting appears at M point.
Whereas, for the free-standing AFM FeSe monolayer, bands $\alpha$
and $\beta$ are symmetric along M-X and M-X$^{\prime}$ directions, and they are
degenerate at M point for both neutral and charged systems as shown
in Fig. \ref{band_charge_doping}. Considering the net charge and
dipole effect of SrTiO$_{3}$ substrate, a vertical electric field
may exist in FeSe/SrTiO$_{3}$. Then we further study the electric
filed effect. In our further DFT calculations, a free-standing monolayer FeSe is particularly designed to be exposed
in a perpendicular electric filed, and the atomic positions are
relaxed for each electric field strength. The calculation results
are shown in Fig. \ref{Efield_DFT}. Clearly, when the electric field
strength is zero, the energy bands are symmetric along M-X and
M-X$^{\prime}$ directions, and the bands of spin-up and spin-down
components are the same. When the electric field strength is nonzero, however, we are astonished to see that
the energy bands along M-X and M-X$^{\prime}$
directions become nonsymmetric, with an obvious spin splitting at M point.
The splitting amplitude increases with increasing
the electric field strength. So we verify that the energy bands near the Fermi
energy and M point are distorted by the perpendicular electric
field, and the energy of band $\alpha$ at X and X$^{\prime}$ points
are different. Besides that, the features of spin-up and spin-down
bands are just opposite, namely, the energy bands in M-X$^{\prime}$
(M-X) direction for spin-up electrons are just the same as the
energy bands in M-X (M-X$^{\prime}$) direction for spin-down
electrons. Comparing the spin-up bands in Fig. \ref{Efield_DFT} for free-standing FeSe monolayer and
those for FeSe monolayer on SiTiO$_{3}$ surface (Fig.
\ref{band}), we see that the distortion of the energy bands are
quite similar. Thus, the effect of self-established electric field
in FeSe/SiTiO$_{3}$ is confirmed.

\begin{figure}[ptb]
\begin{center}
\includegraphics[width=0.5\linewidth]{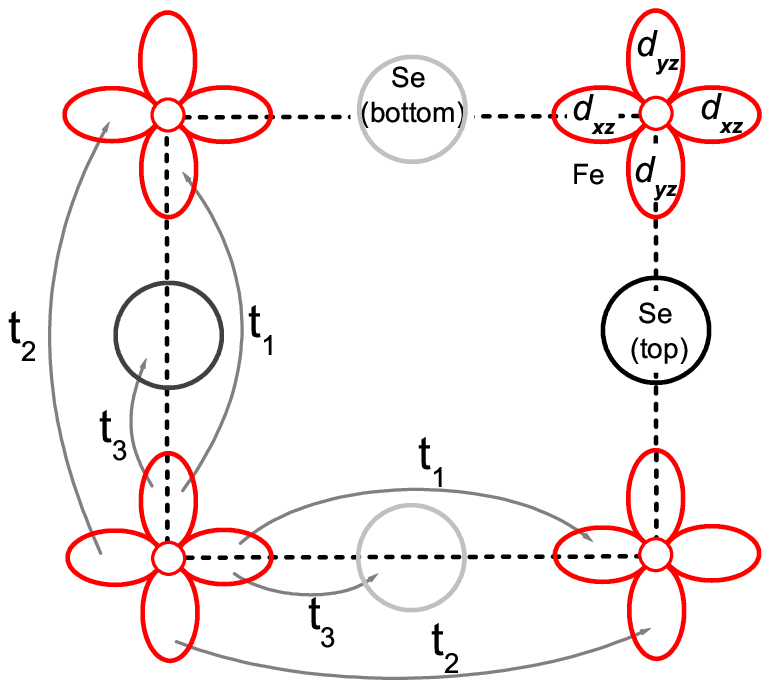}
\end{center}
\caption{(Color online) The hopping schematic diagram of spin-up
tight-binding model in a unit cell. The small red circles, black
large circles and gray large circles show the positions of Fe atoms,
top Se atoms and bottom Se atoms separately. The spin-down Fe atom
is located at the cell center and is not plotted in this spin-up
tight-binding model. The spin-down tight-binding model can be
obtained by exchanging the positions of black and gray circles.}%
\label{TB_model}%
\end{figure}

To explore the detailed mechanism of energy band distortion, we
construct a tight-binding model Hamiltonian to describe the system.
As we have discussed above, the corresponding wavefunctions of bands
$\alpha$ and $\beta$ near M point are composed of $d_{xz}$ and
$d_{yz}$ orbitals of Fe atom, and the wavefunction of energy band
$\beta$ near $\Gamma$ point mainly comes from $d_{z^{2}}$ orbital of Fe
atom. From Fig. \ref{Efield_DFT} we know that the energy band
distortion mainly occurs near M point. Thus, we will use $d_{xz}$
and $d_{yz}$ orbitals of Fe atoms in our tight-binding model, and
neglect the $d_{z^{2}}$ orbital for simplicity. However, the
vertical electric field could not affect a model which only consists
of $d_{xz}$ and $d_{yz}$ orbitals of Fe atoms, since all the centers
of these orbitals are in a horizontal plane. Thus some out-plane
orbitals should be included in our model. The most natural choice is
to use the orbitals of top and bottom Se atoms. We assume the Se
orbitals in our model are symmetric in $xy$-plane, which may be Se 4$p_{z}$ orbital or its combination with Se 4$s$ orbital. The following calculation
shows that this assumption well describes the energy bands along M-X
and M-X$^{\prime}$ directions.

In our mean-field tight-binding model of AFM FeSe monolayer, we
describe the spin-up and spin-down electrons separately. For spin-up
electrons, the tight-binding model contains $d_{xz}$ and $d_{yz}$
orbitals of the spin-up Fe atom, besides one symmetric orbital for
each Se atom. Similarly, for spin-down electrons, the tight-binding
model contains $d_{xz}$ and $d_{yz}$ orbitals of the spin-down Fe
atom, and one symmetric orbital for each Se atom. The hopping of
spin-up electrons is schematically shown in Fig. \ref{TB_model}. For
the spin-down electrons, whereas, the hopping schematic diagram can
be obtained by switching the positions of top and bottom Se atoms.

Considering the symmetry of each orbital, the tight-binding model Hamiltonian
for spin-up electrons is written as $H^{(\text{up})}=H_{\text{on-site}%
}^{(\text{up})}+H_{T}^{(\text{up})}$, where%
\begin{equation}
H_{\text{on-site}}^{(\text{up})}=\sum_{\vec{r}}\left[  \epsilon_{d}%
(c_{xz,\vec{r}}^{+}c_{xz,\vec{r}}+c_{yz,\vec{r}}^{+}c_{yz,\vec{r}}%
)+(\epsilon_{s}+dE)c_{top,\vec{r}}^{+}c_{top,\vec{r}}+(\epsilon_{s}%
-dE)c_{bot,\vec{r}}^{+}c_{bot,\vec{r}}\right]  ,
\end{equation}%
\begin{align}
H_{T}^{(\text{up})}  &  =\sum_{\vec{r}}\left[  t_{1}(c_{yz,\vec{r}}%
^{+}c_{yz,\vec{r}+\hat{y}}+c_{xz,\vec{r}}^{+}c_{xz,\vec{r}+\hat{x}}%
)+t_{2}(c_{xz,\vec{r}}^{+}c_{xz,\vec{r}+\hat{y}}+c_{yz,\vec{r}}^{+}%
c_{yz,\vec{r}+\hat{x}})\right. \\
&  \left.  +t_{3}(c_{xz,\vec{r}}^{+}c_{bot,\vec{r}+\hat{x}/2}+c_{yz,\vec{r}%
}^{+}c_{top,\vec{r}+\hat{y}/2})\right]  +h.c..\nonumber
\end{align}
Here, $c_{xz,\vec{r}}^{+}$ and $c_{yz,\vec{r}}^{+}$ are annihilation operators
of Fe $d_{xz}$ and $d_{yz}$ orbitals, while $c_{top,\vec{r}}^{+}$ and
$c_{bot,\vec{r}}^{+}$ are those of top and bottom Se symmetric orbitals,
respectively. For simplicity the
spin indices are omitted. The parameter $d$ is the distance between top (or
bottom) Se atom and Fe plane, and $E$ are the electric field strength.

The tight-binding model Hamiltonian for spin-down electrons is
similar to that of spin-up electrons, except for that the positions
of top and bottom Se atoms are exchanged. Thus, the Hamiltonian for
spin-down electrons is written as
$H^{(\text{down})}=H_{\text{on-site}}^{(\text{down})}+H_{T}^{(\text{down})}$,
where%
\begin{equation}
H_{\text{on-site}}^{(\text{down})}=\sum_{\vec{r}}\left[  \epsilon
_{d}(c_{xz,\vec{r}}^{+}c_{xz,\vec{r}}+c_{yz,\vec{r}}^{+}c_{yz,\vec{r}%
})+(\epsilon_{s}+dE)c_{top,\vec{r}}^{+}c_{top,\vec{r}}+(\epsilon
_{s}-dE)c_{bot,\vec{r}}^{+}c_{bot,\vec{r}}\right]  ,
\end{equation}%
\begin{align}
H_{T}^{(\text{down})}  &  =\sum_{\vec{r}}\left[  t_{1}(c_{yz,\vec{r}}%
^{+}c_{yz,\vec{r}+\hat{y}}+c_{xz,\vec{r}}^{+}c_{xz,\vec{r}+\hat{x}}%
)+t_{2}(c_{xz,\vec{r}}^{+}c_{xz,\vec{r}+\hat{y}}+c_{yz,\vec{r}}^{+}%
c_{yz,\vec{r}+\hat{x}})\right. \\
&  \left.  +t_{3}(c_{xz,\vec{r}}^{+}c_{top,\vec{r}+\hat{x}/2}+c_{yz,\vec{r}%
}^{+}c_{bot,\vec{r}+\hat{y}/2})\right]  +h.c..\nonumber
\end{align}
By performing the Fourier transformation, the model Hamiltonians for spin-up
and spin-down electrons in momentum space are given by%
\begin{equation}
H^{(\text{up})}(\mathbf{k})=\left[
\begin{array}
[c]{cccc}%
\epsilon_{d_{xz}}(\mathbf{k}) & 0 & -it_{3}\sin(k_{y}/2) & 0\\
0 & \epsilon_{d_{yz}}(\mathbf{k}) & 0 & -it_{3}\sin(k_{x}/2)\\
it_{3}\sin(k_{y}/2) & 0 & \epsilon_{\text{top}} & 0\\
0 & it_{3}\sin(k_{x}/2) & 0 & \epsilon_{\text{bot}}%
\end{array}
\right]  ,
\end{equation}%
\begin{equation}
H^{(\text{down})}(\mathbf{k})=\left[
\begin{array}
[c]{cccc}%
\epsilon_{d_{xz}}(\mathbf{k}) & 0 & -it_{3}\sin(k_{x}/2) & 0\\
0 & \epsilon_{d_{yz}}(\mathbf{k}) & 0 & -it_{3}\sin(k_{y}/2)\\
it_{3}\sin(k_{x}/2) & 0 & \epsilon_{\text{top}} & 0\\
0 & it_{3}\sin(k_{y}/2) & 0 & \epsilon_{\text{bot}}%
\end{array}
\right]  ,
\end{equation}
where $\epsilon_{d_{xz}}(\mathbf{k})$=$\epsilon_{d}$+$t_{1}\cos(k_{y})$%
+$t_{2}\cos(k_{x})$, $\epsilon_{d_{yz}}(\mathbf{k})$=$\epsilon_{d}$+$t_{1}%
\cos(k_{x})$+$t_{2}\cos(k_{y})$, $\epsilon_{\text{top}}$=$\epsilon_{s}$+$dE$,
and $\epsilon_{\text{bot}}$=$\epsilon_{s}\mathtt{-}dE$.

\begin{figure}[ptb]
\begin{center}
\includegraphics[width=0.5\linewidth]{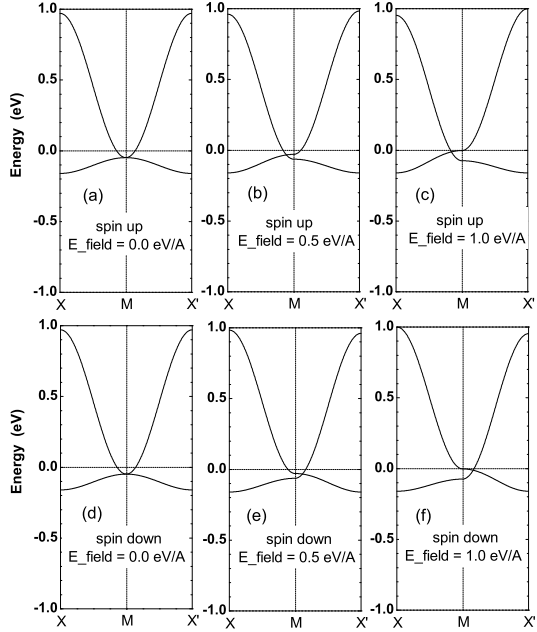}
\end{center}
\caption{(Color online) Calculated tight-binding energy bands of
monolayer FeSe for spin-up (a-c) and spin-down (d-f) channels. The
energy bands distortion is
shown near M point. The Fermi energy is set to zero.}%
\label{TB_band}%
\end{figure}

After diagonalization of Eqs. (5) and (6), the FeSe monolayer's
energy bands near the Fermi surface are plotted in Fig.
\ref{TB_band}, in which we have
used the parameters of $\epsilon_{d}\mathtt{=}$0.36 eV, $\epsilon_{s}%
$=$\mathtt{-}$4.57 eV, $d$=1.37 \r{A}, $t_{1}$\texttt{=}$-$0.01 eV, $t_{2}%
$=0.52 eV and $t_{3}$=0.71 eV. The results show the Fermi surface
distortion clearly. The energy bands are symmetric along X-M-X$^{\prime}$,
and are the same for spin-up and spin-down
electrons. In the presence of a nonzero vertical electric field, an
energy splitting appears at M point. The amplitude of this splitting
increases with increasing the electric field strength. The energy
bands become asymmetric along X-M-X$^{\prime}$. These features are
quantitatively consistent with our DFT results [see Fig.
(\ref{Efield_DFT})]. Thus, our tight-binding model further explains
the origin of energy bands distortion in FeSe/SrTiO$_{3}$. Here, it
should be noticed that in order to describe the electronic
structures for the entire Brillouin zone, the tight-binding model
should contains Fe $d_{z^{2}}$ orbital besides the orbitals
discussed above, since the energy band $\beta$ in Fig. \ref{band} at
$\Gamma$ point is mainly contributed by $d_{z^{2}}$ orbital.

To summarize, by performing the \textit{ab initio} calculations, we have
systematically investigated the electronic structure of AFM FeSe monolayer on
TiO$_{2}$ terminated SrTiO$_{3}$(001) surface. We have shown that the
interaction between SrTiO$_{3}$(001) substrate and AFM FeSe monolayer induces
charge transfer and a self-constructed electric field, which fundamentally
modify the band structure and Fermi surface topology of the FeSe monolayer. In
particular, the energy bands near the Fermi level are dramatically distorted
by the electric field and show an obvious spin-resolved splitting around M
point in BZ. As a result, the spin-resolved electron pockets evolved from
band $\alpha$ become anisotropic and flat. We have proposed a tight-binding
model Hamiltonian to reproduce these features. Also, we have shown that the
present monolayer FeSe/SrTiO$_{3}$(001) composite structure is an ideal
electron-hole bilayer system, with electrons and holes respectively formed in
FeSe monolayer and TiO$_{2}$. These findings are to be verified by the future
spin-polarized ARPES experiments and are expected to shed light on the
high-temperature superconductivity at the oxide interface.

This work was supported by Natural Science Foundation of China under Grants
No. 90921003 and No. 11004013, and by the National Basic Research Program of
China (973 Program) under Grant No. 2009CB929103.

\end{document}